\documentclass[12pt,aps,groupedaddress,superscriptaddress,pra,preprint]{revtex4}

\usepackage{graphicx}
\usepackage{amsmath,amssymb,amsfonts}
\usepackage{natbib}

\begin{document}

\title{Understanding chemical reactions within a generalized
Hamilton-Jacobi framework}

\author{A. S. Sanz}
\email{asanz@imaff.cfmac.csic.es}
\affiliation{Instituto de F\'{\i}sica Fundamental,
Consejo Superior de Investigaciones Cient\'{\i}ficas,
Serrano 123, 28006 Madrid, Spain}

\author{X. Gim\'enez}
\affiliation{Departament de Qu\'{\i}mica F\'{\i}sica, Universitat de Barcelona,
Mart\'{\i} i Franqu\`es 1, 08028 Barcelona, Spain}
\affiliation{Institut de Qu\'{\i}mica Te\`orica i Computacional de la
Universitat de Barcelona, Mart\'{\i} i Franqu\`es 1, 08028 Barcelona, Spain}

\author{J. M. Bofill}
\affiliation{Institut de Qu\'{\i}mica Te\`orica i Computacional de la
Universitat de Barcelona, Mart\'{\i} i Franqu\`es 1, 08028 Barcelona, Spain}
\affiliation{Departament de Qu\'{\i}mica Org\`anica, Universitat de
Barcelona, Mart\'{\i} i Franqu\`es 1, 08028 Barcelona, Spain}

\author{S. Miret-Art\'es}
\affiliation{Instituto de F\'{\i}sica Fundamental,
Consejo Superior de Investigaciones Cient\'{\i}ficas,
Serrano 123, 28006 Madrid, Spain}

\date{\today}

\begin{abstract}
Reaction paths and classical and quantum trajectories are studied
within a generalized Hamilton-Jacobi framework, which allows to put
on equal footing topology and dynamics in chemical reactivity problems.
In doing so, we show how high-dimensional problems could be dealt
with by means of Carath\'eodory plots or how trajectory-based
quantum-classical analyses reveal unexpected discrepancies.
As a working model, we consider the reaction dynamics associated with
a M\"uller-Brown potential energy surface, where we focus on the
relationship between reaction paths and trajectories as well as
on reaction probability calculations from classical and quantum
trajectories.
\end{abstract}

\pacs{}

\maketitle

%%%%%%%%%%%%%%%%%%%%%%%%%%%%%%%%%%%%%%%%%%%%%%%%%%%%%%%%%%%%%%%%%%%%%%%

\section{Introduction}
\label{sec1}

One of the most outstanding problems in theoretical Chemical Physics
is studying and determining the mechanisms that underly chemical
reactions \cite{faradaydisc}.
Traditionally, a full theoretical study of chemical reactions involves
two steps: (1) determining a potential energy surface (PES) and (2) the
analysis of the associated dynamics.
This has given rise to two types of comparative schemes.
Within the first scheme, comparisons are established between the
topological properties of PESs and the dynamics ruled by them.
Because most of the energy exchange is carried by electrons in the
reactants-to-products transition, chemical reactions are usually studied
neglecting the (tinny) contribution of the nuclei energy.
This has led to the `static', topological analysis of chemical
reactions, and also to an endless debate between two communities:
PES explorers and dynamicists.
The second scheme is based on quantum-classical correspondence, i.e.,
on comparing the differences displayed by an observable when computed
classically and quantum-mechanically.
Any deviation between these two calculations is then regarded as a
signature of `quantumness', this being a standard criterion to
discern whether the observable is affected or not by quantum effects.
This argument, however, can be misleading: the classical and quantum
values could be quite similar, but the dynamics leading to them could
be very different.
Therefore, analyzing quantum effects in terms of individual events
(trajectories) would be highly desirable, which can be achieved within
a theoretical framework based on the Hamilton-Jacobi (HJ) formalism.

In the analysis of PESs, molecular processes are described by means of
{\it reaction paths} (RPs), which are continuous curves {\it on} the
PES connecting two minima through a saddle point (first-order maximum)
and usually associated with {\it steepest-descent curves} \cite{fukui}.
The minima are usually related to the reactants and products states,
while the saddle point describes the reaction transition state.
All these elements provide us with a mechanistic description where the
chemical reaction consists of a sequence of RP points, each one related
to a particular molecular geometry.
However, molecular geometry changes as the reaction evolves in time
cannot be explained satisfactorily, what has derived in dynamical
approaches and, therefore, the first type of comparative scheme.
In this regard, former connections between the RP formalism and the
classical HJ equation can be found in \cite{bofill1}, for instance.

The second comparison scheme arises when we realize that the fundamental
parameters characterizing chemical systems (masses, energies, density
distributions, etc.) very often lie on the border between purely
classical and quantum behaviors.
Actually, many times, classifying such behaviors results a hard task
and one has to resort to previous experience before labelling them as
classical, quantum or `in between'.
In this sense, both formal \cite{vleck,schiller1,schiller2} and
computational analyses can be found frequently in the literature,
the latter usually considered simultaneously with the former
\cite{miller1,miller2}.
The problem is that the solutions obtained from classical and
quantum evolution equations are formally and conceptually different.
The classical output is a trajectory, a time-ordered list of positions
and momenta; the quantum output, however, is a wave function and,
therefore, a time-ordered list of probabilities covering the whole
position {\it or} momentum space.
Thus, before comparing classical and quantum outputs, some averaging
of the former is required in order to approach the statistical nature
of the latter (semiclassical mechanics constitutes a mixed approach,
where a classical-like view of quantum effects is achieved after
some quantization scheme is carried out).
The calculation of reaction probabilities and/or cross sections in
molecular collisions, or Franck-Condon factors in molecular spectra
are well-known examples.
Though classical-trajectory-based studies (and methodologies) are very
common within the molecular reaction literature, similar quantum
trajectory treatments can also be found but scarcely.
For instance, Muga et al.~\cite{muga1,muga2} have computed quantum
moments from an averaging procedure and then compared with classical
quantities.
Within the framework of Quantum Hydrodynamics \cite{Madelung},
i.e., treating the probability density as a quantum fluid, chemical
reactions were formerly studied by McCullough and Wyatt
\cite{MC-W1,MC-W2,MC-W3}.
Similarly, in Bohmian mechanics \cite{bohm,holland}, the evolution of
the system is described in terms of quantum trajectories, which can be
obtained through any of two approaches \cite{wyatt-bk}: synthetic
\cite{wyatt-bk,wyatt1,wyatt2} and analytic
\cite{sanz-ssr,sanz-pr,bofill2}.
Within the former, quantum trajectories are computed after solving
simultaneously the quantum HJ equation and the continuity one, while in
the latter (used in this work) one starts with Schr\"odinger's equation
and then the trajectories are obtained from the phase of the wave
function.

In this work, we explore and discuss the connection between topology and
dynamics in chemical reactions within a generalized HJ framework.
This general scenario is introduced in Sec.~\ref{sec2}.
Though the corresponding equations are conceptually different, their
underlying common formal structure allows us to establish a connection
among them.
This is illustrated in Sec.~\ref{sec3}, where we study the dynamics
associated with the M\"uller-Brown PES model \cite{muellerbrown,gb-pccp},
which can be used to simulate, for example, the passage from reactants
to products in isomerization or enzymatic (e.g., Michaelis-Menten)
reactions \cite{isaacs}.
Classical and quantum reaction probabilities are calculated from the
corresponding trajectory statistics.
Then, a comparative analysis between classical and quantum trajectories
is presented, as well as the relationship between the latter and the RP
through Carath\'eodory plots \cite{cara}.
Finally, the main conclusions from this work are summarized in
Sec.~\ref{sec4}.

%%%%%%%%%%%%%%%%%%%%%%%%%%%%%%%%%%%%%%%%%%%%%%%%%%%%%%%%%%%%%%%%%%%%%%%

\section{A common theoretical framework to analyze chemical reactions}
\label{sec2}

The question of whether there exists a common theoretical ground to
the geometric (topological), classical and quantum descriptions of
chemical reactions emerges naturally when comparing the steps (1) and
(2) above.
A priori, they imply two distant mathematical frameworks
\cite{faradaydisc}: the PES description involves analyses strictly
based on topology, while the dynamical ones (classical or
quantum-mechanical) are grounded on the HJ formalism.
However, a closer inspection shows that both frameworks are not so
far one from another.
As shown in the literature \cite{bofill}, within the so-called
{\it reaction path models} \cite{heidrich,schlegel}, PESs can be
described by an expression similar to the classical HJ equation.
Now, based on the fact that both the PES topology and classical
dynamics are endowed with a HJ equation, it was suggested long time
ago \cite{courant} the possibility to associate a wave propagation
within the HJ formalism, this establishing a link with quantum
mechanics. Therefore, it should be possible to understand the three
frameworks on similar grounds, in particular, through the so-called
{\it Huyghens construction} \cite{courant}, very well known in Optics
\cite{bornwolf}, for instance.
This construction is a geometric method for mapping the progress in
terms of some parameter of surfaces of equal phase, which satisfy at
each point a HJ-like equation.
RPs in the topological case and trajectories in the dynamical one
then constitute the set of solutions (characteristics) crossing
transversally a family of such surfaces with the same phase (or
equidistant).

To understand the analogy between RPs and trajectories, note that both
can be obtained from similar variational principles.
That is, consider the functional
\begin{equation}
 I[{\bf x}(\tau)] =
  \int_{\tau_0}^\tau F({\bf x}, \dot{\bf x}, \tau') d\tau' ,
 \label{eq-1}
\end{equation}
where ${\bf x} = (x_1, x_2, \ldots, x_n)$ denotes a coordinate vector
and $\dot{\bf x} = d{\bf x}/d\tau$ its derivative with respect to an
independent parameter $\tau$ (physically, this parameter is related to
the reaction coordinate, $s$, in the RP formalism and to the time,
$t$, in the dynamical ones).
RPs and trajectories are curves ${\bf x}(\tau)$ for which
the integral (\ref{eq-1}) becomes an extremum (maximum or minimum)
under small variations of the coordinates (i.e., $\delta I = 0$).
This gives rise to the well-known Euler-Lagrange equations,
\begin{equation}
 \frac{\partial F}{\partial x_i}
  - \frac{d}{d\tau} \frac{\partial F}{\partial \dot{x}_i} = 0 ,
 \qquad i = 1, 2, \ldots, n ,
 \label{eq-2}
\end{equation}
which are satisfied when $F({\bf x}, \dot{\bf x}, \tau)$ is evaluated
along the solution ${\bf x}(\tau)$.
Alternatively, by means of Legendre transformations, it can also be
shown \cite{courant} that any of these solutions is also a solution
of a HJ-like equation,
\begin{equation}
 G({\bf x},\nabla_{\bf x}I,\partial I/\partial \tau, \tau) \equiv
  \frac{\partial I}{\partial \tau}
  + \frac{1}{2}\ \! \nabla_{\bf x}^T I \cdot \nabla_{\bf x} I
  + \mathcal{V}({\bf x},\tau) = 0 ,
 \label{eq-3}
\end{equation}
where $\mathcal{V}({\bf x},t)$ is a general potential function.
The intrinsic structure of $F({\bf x}, \dot{\bf x}, \tau)$ is different
within each one of the three frameworks.
However, some similar elements (characteristic equations) can still be
found among them, as seen below.

In the RP formalism, the associated $F$ is a homogeneous functional of
degree one with respect to $\dot{\bf x}$ \cite{bofill1}, i.e.,
\begin{equation}
 F({\bf x}, \dot{\bf x}, s)
  = [{\bf g}^T({\bf x}) \cdot {\bf g}({\bf x})]^{1/2}
      [\dot{\bf x}^T \cdot \dot{\bf x}]^{1/2} ,
 \label{eq-4}
\end{equation}
where ${\bf g}({\bf x})$ is the PES gradient at ${\bf x}$ \cite{quapp}
and the superscript $T$ denotes the transpose vector.
The corresponding HJ-like equation (\ref{eq-3}) can then be expressed
\cite{bofill1} as
\begin{equation}
  \frac{1}{2}\ \! \nabla_{\bf x}^T V({\bf x}) \cdot
     \nabla_{\bf x} V({\bf x})
  - \frac{1}{2}\ \! {\bf g}^T({\bf x}) \cdot {\bf g}({\bf x}) = 0 ,
 \label{eq-5}
\end{equation}
where the term $\partial I/\partial s$ is lacking because the PES
(here, playing the role of the surface $I$) does not depend explicitly
on $s$, but only on ${\bf x}$.
From (\ref{eq-5}),
\begin{equation}
 {\bf g}({\bf x}) = \nabla_{\bf x} V({\bf x}) ,
 \label{eq-6}
\end{equation}
which, after integration over $s$, gives the steepest-descent curve,
${\bf x}(s)$, describing the RP, the curve joining two PES minima
through a saddle point.
This equation is analogous to that found in Optics to determine the
optical path followed by light in media with variable refraction
indexes, according to Fermat's principle \cite{bornwolf}, or the
geodesic equation in Gravitation \cite{landau}.

In classical dynamics, Fermat's principle translates into Hamilton's
principle, and the integrand of (\ref{eq-1}) is just the Lagrangian
function \cite{goldstein}, which reads as
\begin{equation}
 F({\bf x}, \dot{\bf x}, t) =
  \frac{1}{2}\ \! \dot{\bf x}^T \cdot \dot{\bf x} - V({\bf x})
 \label{eq-7}
\end{equation}
in reduced length units (i.e., ${\bf x}/\sqrt{m} \to {\bf x}$).
The corresponding HJ equation (\ref{eq-3}) is
\begin{equation}
 \frac{\partial S_{\rm cl}({\bf x},t)}{\partial t}
  + \frac{1}{2}\ \! \nabla_{\bf x}^T S_{\rm cl}({\bf x},t) \cdot
       \nabla_{\bf x} S_{\rm cl}({\bf x},t)
  + V({\bf x}) = 0 ,
 \label{eq-8}
\end{equation}
where its solution, the classical action $S_{\rm cl}$, follows the
Huyghens construction \cite{courant}.
Usually, in the kind of process that we are interested in, the total
energy $E$ conserves and, therefore, $-\partial S_{\rm cl}/\partial t
= E$.
This allows to reexpress (\ref{eq-8}) as
\begin{equation}
  \frac{1}{2}\ \! \nabla_{\bf x}^T S_{\rm cl}({\bf x},t) \cdot
    \nabla_{\bf x} S_{\rm cl}({\bf x},t)
  - [E - V({\bf x})] = 0 ,
 \label{eq-9}
\end{equation}
which has the same formal structure as Eq.~(\ref{eq-5}).
Similarly to RPs, classical trajectories are now obtained from
\begin{equation}
 \dot{\bf x} = {\bf p}(t) = \nabla_{\bf x} S_{\rm cl}({\bf x},t)
 \label{eq-10}
\end{equation}
by integrating over time with initial conditions ${\bf x}_0$ and
${\bf p}_0$.
Note that, from this relation and Eq.~(\ref{eq-9}), we can readily
derive the well-known Newtonian expression for the velocity
\begin{equation}
 |\dot{\bf x}| = \sqrt{2[E - V({\bf x})]} .
 \label{eq-11}
\end{equation}

Unlike the steepest-descent curve, given a PES there is an infinite
number of associated classical trajectories, as many as initial
${\bf p}_0$ one can provide for a given ${\bf x}_0$ and $E$.
Thus, in chemical reactivity problems, a single trajectory results
meaningless by itself; in order to extract valuable information about
the process, a distribution of them (i.e., a sampling over initial
conditions) has to be considered.
This statistical problem, equivalent to consider an ensemble of
identical, non-interacting systems described initially by some
pre-assigned density distribution, $\rho_{\rm cl} ({\bf x},0)$,
can be expressed in terms of a Lagrangian density \cite{holland} as
\begin{equation}
 F(\rho_{\rm cl}, \nabla_{\bf x} \rho_{\rm cl}, \dot{\rho}_{\rm cl},
  S_{\rm cl}, \nabla_{\bf x} S_{\rm cl}, \dot{S}_{\rm cl}; {\bf x}, t)
 = - \left[ \frac{1}{2}\ \! \nabla^T S_{\rm cl}({\bf x},t) \cdot
   \nabla S_{\rm cl}({\bf x},t)
   - V({\bf x}) \right] \rho_{\rm cl}({\bf x},t) ,
 \label{eq-12}
\end{equation}
where the variables are now the fields $\rho_{\rm cl}$ and $S_{\rm cl}$,
and the independent parameters are ${\bf x}$ and $t$.
Applying the Euler-Lagrange equation to (\ref{eq-12}) with respect to
the field variables yields (\ref{eq-8}) and
\begin{equation}
 \frac{\partial \rho_{\rm cl}({\bf x},t)}{\partial t} +
  \nabla_{\bf x}^T \cdot [\rho_{\rm cl}({\bf x},t)
    \nabla_{\bf x} S({\bf x},t)] = 0 ,
 \label{eq-13b}
\end{equation}
which is the classical Liouville equation describing a swarm of single,
non-interacting particles each one evolving according to (\ref{eq-8}).
From (\ref{eq-13b}), note that the time evolution of trajectories is
totally independent of the density distribution evolution.
This can be inferred from Eq.~(\ref{eq-12}), which is separable in
$S_{\rm cl}$ and $\rho_{\rm cl}$.
The only relationship between $\rho_{\rm cl}$ and individual
trajectories is that the former is used to choose initial conditions
for the latter, this being a statistical rather than a dynamical
relation.

It is known \cite{holland,goldstein} that the wave formulation of
quantum mechanics can be derived from a Lagrangian density
\begin{eqnarray}
 F(\Psi, \nabla_{\bf x} \Psi, \dot{\Psi},
  \Psi^*, \nabla_{\bf x} \Psi^*, \dot{\Psi}^*; {\bf x}, t) & = &
  \frac{i\hbar}{2} \left[ \Psi^* ({\bf x},t)\ \!
   \frac{\partial \Psi ({\bf x},t)}{\partial t}
  - \frac{\partial \Psi^* ({\bf x},t)}{\partial t} \ \!
     \Psi ({\bf x},t) \right] \nonumber \\
 & - & \frac{\hbar^2}{2}\ \! \nabla_{\bf x}^T \Psi ({\bf x},t) \cdot
      \nabla_{\bf x} \Psi^* ({\bf x},t)
  - V({\bf x}) \Psi ({\bf x},t) \Psi^* ({\bf x},t)
 \nonumber \\ & &
 \label{eq-14}
\end{eqnarray}
when we require the corresponding integral to be stationary with
respect to variations in the complex-valued field variables $\Psi$
and $\Psi^*$.
By applying the Euler-Lagrange equation with respect to $\Psi$, we
obtain the time-dependent Schr\"odinger equation,
\begin{equation}
 i \hbar \ \! \frac{\partial \Psi ({\bf x},t)}{\partial t} =
  - \frac{\hbar^2}{2} \ \! \nabla_{\bf x}^2 \Psi ({\bf x},t)
  + V({\bf x}) \Psi ({\bf x},t)
 \label{eq-15}
\end{equation}
(and its conjugate complex if $\Psi^*$ is considered), where
$\nabla_{\bf x}^2 \equiv \nabla_{\bf x}^T \cdot \nabla_{\bf x}$.
However, there is an alternative way to proceed consisting of switching
from $\Psi$ and $\Psi^*$ to the real-valued field variables $S$ and
$\rho$ according to the transformation \cite{bohm,takabayasi}:
$\Psi({\bf x},t) = \rho^{1/2}({\bf x},t) e^{i S({\bf x},t)/\hbar}$.
Substituting this expression into (\ref{eq-14}) readily yields the new
Lagrangian density \cite{takabayasi},
\begin{eqnarray}
 F(\rho, \nabla_{\bf x} \rho, \dot{\rho}, S, \nabla_{\bf x} S, \dot{S};
  {\bf x}, t)
 & = & - \left[ \frac{1}{2}\ \! \nabla_{\bf x}^T S({\bf x},t) \cdot
   \nabla_{\bf x} S({\bf x},t) - V({\bf x}) \right] \rho({\bf x},t)
 \nonumber \\
 & & - \frac{\hbar^2}{8} \frac{\nabla_{\bf x}^T \rho({\bf x},t) \cdot
   \nabla_{\bf x} \rho({\bf x},t)}{\rho({\bf x},t)} .
 \label{eq-17}
\end{eqnarray}
Now, after inserting $F$ into the Euler-Lagrange equation, we obtain
\begin{subequations}
 \begin{eqnarray}
 \frac{\partial S({\bf x},t)}{\partial t} & + &
 \frac{1}{2}\ \! \nabla_{\bf x}^T S({\bf x},t) \cdot
    \nabla_{\bf x} S({\bf x},t)
  + V_{\rm eff}({\bf x},t) = 0 ,
 \label{eq-18a} \\
 \frac{\partial \rho ({\bf x},t)}{\partial t} & + &
  \nabla_{\bf x}^T \cdot [\rho ({\bf x},t)
    \nabla_{\bf x} S({\bf x},t)] = 0 ,
\label{eq-18b}
 \end{eqnarray}
\label{eq-18}
\end{subequations}
where the {\it effective potential}
\begin{equation}
 V_{\rm eff} ({\bf x},t) = V({\bf x}) - \frac{\hbar^2}{4}
  \left[ \frac{1}{2}
  \frac{\nabla_{\bf x}^T \rho({\bf x},t)}{\rho({\bf x},t)} \cdot
  \frac{\nabla_{\bf x} \rho({\bf x},t)}{\rho({\bf x},t)}
    - \frac{\nabla_{\bf x}^2 \rho({\bf x},t)}{\rho({\bf x},t)}\right]
 \label{eq-19}
\end{equation}
is a sum of the PES and the so-called {\it quantum potential}
\cite{bohm}.
Equations (\ref{eq-18a}) and (b) are, respectively, the quantum HJ
equation and the quantum Liouville equation, which are coupled, this
being the origin of quantum effects.
Note that, unlike their classical analogs, quantum trajectories undergo
a (nonlocal) dependence on their distribution: though (\ref{eq-18b})
describes the evolution of an ensemble of non-interacting particles
(there is no physical potential, like $V$, acting among them), the
quantum potential mediates a sort of information exchange among them
which strongly determines their dynamical evolution.
Similarly to the classical case, quantum trajectories are obtained
from
\begin{equation}
 \dot{\bf x} = {\bf p}(t) = \nabla_{\bf x} S({\bf x},t) ,
 \label{eq-20}
\end{equation}
with initial conditions ${\bf x}_0$, $S({\bf x},0)$ and
$\rho({\bf x},0)$; here, the initial momentum is predetermined by the
initial phase of the wave function and, therefore, unlike classical
trajectories, we cannot choose it arbitrarily.
Although $S$ is a multivalued function (i.e., $S'({\bf x},t) =
S({\bf x},t) + 2\pi n \hbar$, with $n$ being an integer), this does
not affect the calculation of trajectories, since only its gradient
is needed.
Only when $\rho$ vanishes, this property plays a fundamental role, for
it rules the appearance of vorticality \cite{holland,sanz-ssr,sanz-pr}.
Moreover, as before, $S({\bf x},t)$ and its gradient vector field,
$\nabla_{\bf x} S({\bf x},t)$, constitute the basic elements for the
Huyghens construction.

\begin{figure}
 \includegraphics[width=22cm,angle=90]{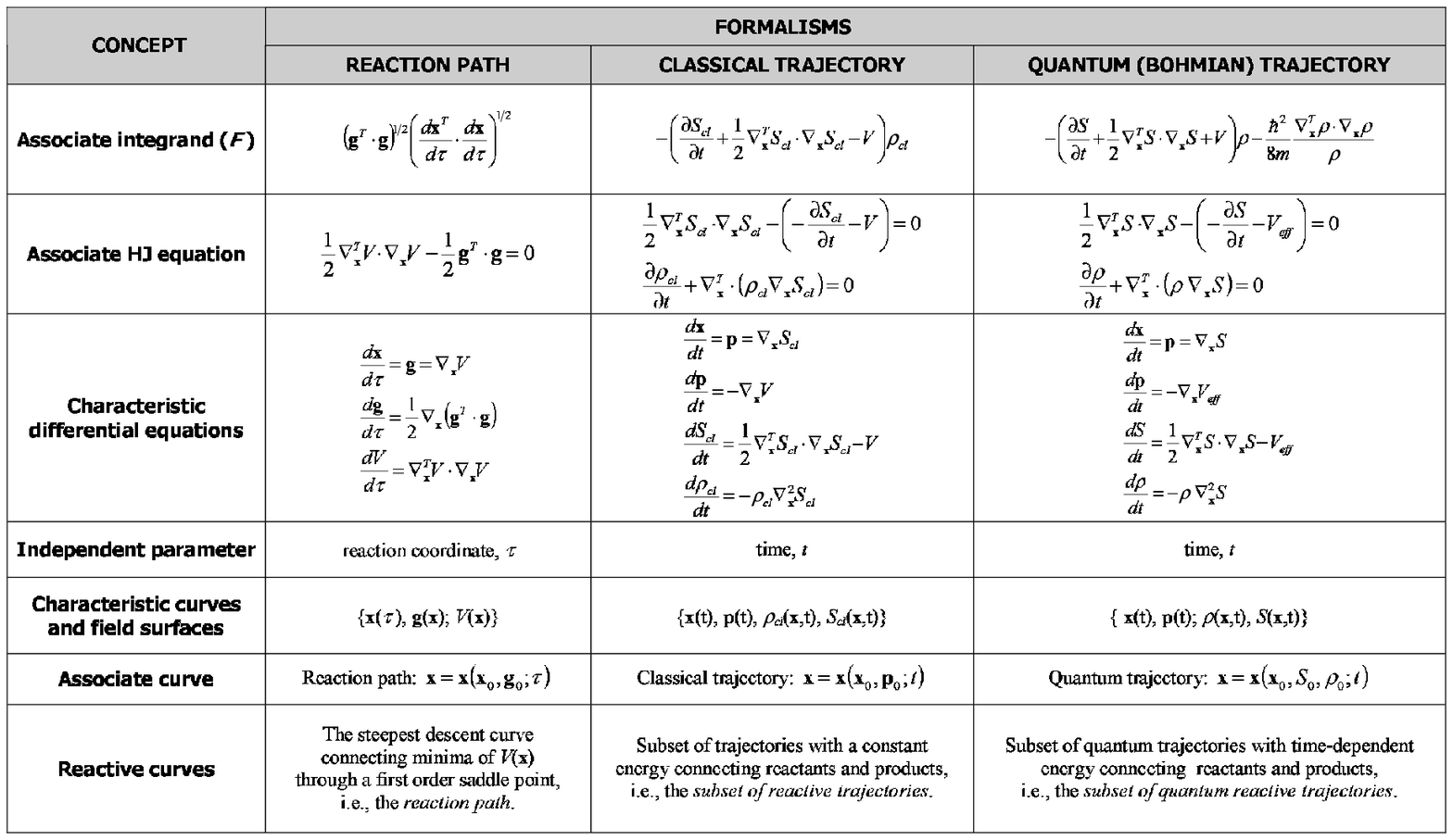}
 \caption{\label{fig1}
  A schematic view of the generalized HJ framework presented in this
  work.}
\end{figure}

The generalized formulation discussed here is summarized in
Fig.~\ref{fig1}.
As seen, mathematically RPs and classical/quantum trajectories are both
characteristics coming from similar differential equations.
However, important differences arise when the terms of the
corresponding equations are identified.
Thus, for a given PES, the RP follows a unique geodesic,
steepest-descent curve, while there is an infinite set of
classical/quantum trajectories solutions of the same differential
equation (\ref{eq-10})/(\ref{eq-20}), each one for an initial
condition.
Nonetheless, there is a common point: if one computes a set of
steepest-descent curves for some given initial conditions ${\bf x}_0$
and ${\bf g}_0$, one fully `builds' the PES.
Similarly, from the set of classical trajectories for some given
initial conditions $E$ and ${\bf x}_0$, one will then `build' the
action hypersurface $S_{cl}({\bf x},t)$ (the same for quantum
trajectories).
Formally, all the characteristics are obtained as solutions of
(\ref{eq-3}), i.e., Eqs.~(\ref{eq-6}), (\ref{eq-10}) and (\ref{eq-20}),
which present similar functional forms.
However, while $\tau$ represents the RP reaction coordinate and for
each $\tau$ we have different values of the PES, it plays the role of
time for a dynamical (classical or quantum) trajectory, which means
that at every time we have a different value of the action
hypersurface.

%%%%%%%%%%%%%%%%%%%%%%%%%%%%%%%%%%%%%%%%%%%%%%%%%%%%%%%%%%%%%%%%%%%%%%%

\section{Numerical simulations}
\label{sec3}

%%%%%%%%%%%%%%%%%%%%%%%%%%%%%%%%%%%%%%%%%%%%%%%%%%%%%%%%%%%%%%%%%%%%%%%

\subsection{The model}
\label{sec3.1}

\begin{figure}
\includegraphics[width=7cm]{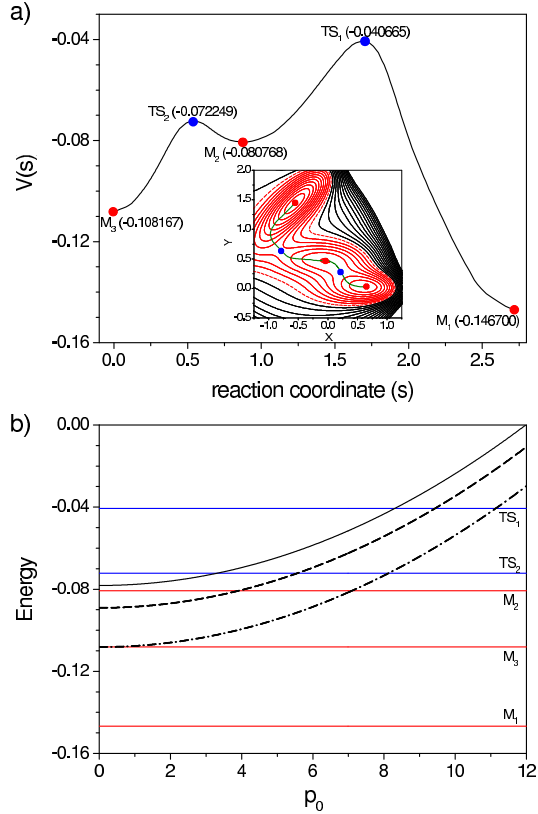}
 \caption{\label{fig2}
 (a) Transition state energies ($TS$; blue circles) and reactants,
 pre-equilibrium and products energies ($M$; red circles) along the
 RP for the M\"uller-Brown PES.
 In the inset, a contour-plot of this PES with the RP (green line);
 black/red contours represent positive/negative equipotential contours.
 (b) Energy diagram as a function of the initial momentum (see text
 for details).}
\end{figure}

As a working model, we consider the M\"uller-Brown PES
\cite{muellerbrown}, which describes the reactants-to-products
passage through an intermediate (pre-equilibrium) state,
\begin{equation}
 R \rightleftharpoons I \rightharpoonup P .
 \label{eq-21}
\end{equation}
This PES has three minima, $M_1(-0.558,1.442)$, $M_2(-0.050,0.467)$
and $M_3(0.623,0.028)$ (within this work, all magnitudes are given in
atomic units), which describe the products, intermediate and reactants
states, respectively.
It has also two transition states, $TS_1 (-0.822,0.624)$ and $TS_2
(0.212,0.293)$, which separate products from pre-equilibrium and the
latter from reactants, respectively.
All these energies are indicated in Fig.~\ref{fig2}a along the RP,
which is described in terms of the arc-length
\begin{equation}
 s(x,y) \approx \sum_{i=1}^N \sqrt{\Delta x_i^2 + \Delta y_i^2}
  = \sum_{i=1}^N \sqrt{(x_i - x_{i-1})^2 + (y_i - y_{i-1})^2} ,
 \label{eq-23}
\end{equation}
where $(x_0,y_0) = M_3$, $(x_N,y_N) = (x,y)$, and the final point of
the RP is $M_1$.
Red circles represent the energies associated with minima ($V_{M_1} =
-0.147$, $V_{M_2} = -0.081$ and $V_{M_3} = -0.108$) and blue ones with
maxima ($V_{TS_1} = -0.041$ and $V_{TS_2} = - 0.072$).
A contour-plot of the M\"uller-Brown PES with the RP (green line) is
displayed in the inset.

Regarding the dynamics, we have considered an initial Gaussian wave
packet,
\begin{equation}
 \Psi_0(x,y) = A_0 \ \! e^{-(x-x_0)^2/4\sigma_x^2
  -(y-y_0)^2/4\sigma_y^2 + i p_{x,0} (x - x_0)/\hbar
  + i p_{y,0} (y - y_0)/\hbar} ,
 \label{eq-24}
\end{equation}
where $A_0 = (2\pi \sigma_x \sigma_y)^{-1/2}$, with $\sigma_x^2 =
\sigma_y^2 = \sigma_0^2 = 0.0125$. This wave packet describes a
proton transfer process and, hence, $m = 1,836$ in the Schr\"odinger
and Bohmian motion equations.
In all calculations, the initial position of the wave-packet center is
kept fixed at $(x_0,y_0) = M_3$, and only its initial momentum, chosen
in all cases as $(p_{x,0},p_{y,0}) = (-p_0,p_0)$, has been varied
(indeed, the value of $p_0$).
The initial conditions for the quantum trajectories are obtained
randomly by sampling $\rho_0 = |\Psi_0|^2 \delta(p_x - p_{x,0}) \delta
(p_y - p_{y,0})$, while for classical trajectories we consider this
distribution as well as the Wigner one associated with (\ref{eq-24}),
\begin{equation}
 \rho_W(x,p_x;y,p_y) \propto
  e^{-(x-x_0)^2/2\sigma_x^2 - \sigma_x^2 (p_x-p_{x,0})^2/\hbar^2
     -(y-y_0)^2/2\sigma_y^2 - \sigma_y^2 (p_y-p_{y,0})^2/\hbar^2} .
 \label{eq-32}
\end{equation}
Fig.~\ref{fig2}b shows an energy diagram as a function of $p_0$,
useful to understand the role played by the choice of initial momenta
in classical/quantum trajectory simulations and, therefore, in the
calculation of reaction probabilities (see Sec.~\ref{3.2}).
In this diagram, the energy of the transition states and minima are
indicated by color horizontal lines (blue and red, respectively).
The black solid line represents the quantum expectation or average
value of the energy, $\bar{E} = \langle \hat{H} \rangle$, which
coincides with the classical average energy when the (classical)
initial positions and momenta are chosen according to (\ref{eq-32}),
since
\begin{equation}
 \bar{E} = \int E(x,p_x;y,p_y) \rho_W(x,p_x;y,p_y) dx dy dp_x dp_y
  = \frac{p_0^2}{m} + \bar{V} + \bar{\delta}
     \approx \frac{1}{N} \sum_{i=1}^N E_i (x_0^i,y_0^i) ,
 \label{eq-33}
\end{equation}
where $\bar{V} = \sum_{i=1}^N V_i(x_0^i,y_0^i)/N$, $\bar{\delta} =
\hbar^2/4m\sigma^2$ is the {\it spreading ratio} \cite{sanz-jpa}, and
the last sum runs over all (classical/quantum) particles considered
(with ($x_0^i, y_0^i$) denoting their corresponding initial positions).
The black dashed line is (\ref{eq-33}), but without $\bar{\delta}$,
which coincides with a classical ensemble distributed initially
according to $\rho_0$.
In classical mechanics, the spreading ratio can be, therefore, related
to the initial momentum distribution around $(p_{x,0},p_{y,0})$.
Finally, the black dash-dotted line gives the energy of a classical
particle initially located at $M_3$ and with initial momentum
$(-p_0,p_0)$.

%%%%%%%%%%%%%%%%%%%%%%%%%%%%%%%%%%%%%%%%%%%%%%%%%%%%%%%%%%%%%%%%%%%%%%%

\subsection{Trajectory ensemble analysis of reaction probabilities}
\label{3.2}

The analysis of reaction probabilities can be carried out by defining
the reaction probability in terms of the restricted norm
\cite{sanz-jcp-sars}
\begin{equation}
 \mathcal{P}(t) \equiv \int_\Sigma |\Psi(x,y,t)|^2 dx dy ,
 \label{eq-28}
\end{equation}
where $\Sigma$ denotes the space region above the line $y_{R \to P}(x)
= 0.8024 x + 1.2734$, here chosen as the frontier line separating
products from the pre-equilibrium/reactants region (more refined
choices could be considered, but no significant discrepancies are
expected regarding products formation at this level).
Apart from the amount of products in time, the initial slope of this
function also provides information about the reaction velocity.

From a Bohmian viewpoint, (\ref{eq-28}) reads as the number of
trajectories $N_\Sigma$ penetrating into $\Sigma$ at a time $t$ with
respect to the total number $N$ of trajectories initially considered
\cite{sanz-jcp-sars},
\begin{equation}
 \mathcal{W}(t) \equiv \frac{N_\Sigma (t)}{N} ,
 \label{eq-29}
\end{equation}
which approaches $\mathcal{P}(t)$ when $N \to \infty$ and the
initial conditions are sampled according to $\rho_0$.
Classically, (\ref{eq-29}) can also be applied, but
$\mathcal{W}_{\rm cl}(t)$ meaning a classical products
fraction.
Probability can flow backwards from products to reactants \cite{MC-W1},
which may lead to a multiple crossing of $y_{R \to P}(x)$ by the same
quantum/classical trajectory.
Hence, another interesting quantity is the fraction of trajectories
going from reactants to products without taking into account their
return to reactants,
\begin{equation}
 \bar{\mathcal{W}}(t) \equiv \frac{\bar{N}_\Sigma (t)}{N} ,
 \label{eq-30}
\end{equation}
which, at $t \to \infty$, provides the total amount of products for a
given initial state.
That is, assuming that one could extract the products formed during the
reaction by some chemical or physical procedure, $\bar{\mathcal{W}}(t)$
would provide the maximum amount of products.
Note that this information is only available when working with
trajectories, since one can visualize each individual process and,
therefore, detect when one reactive event has taken place.

\begin{figure}
\includegraphics[width=7cm]{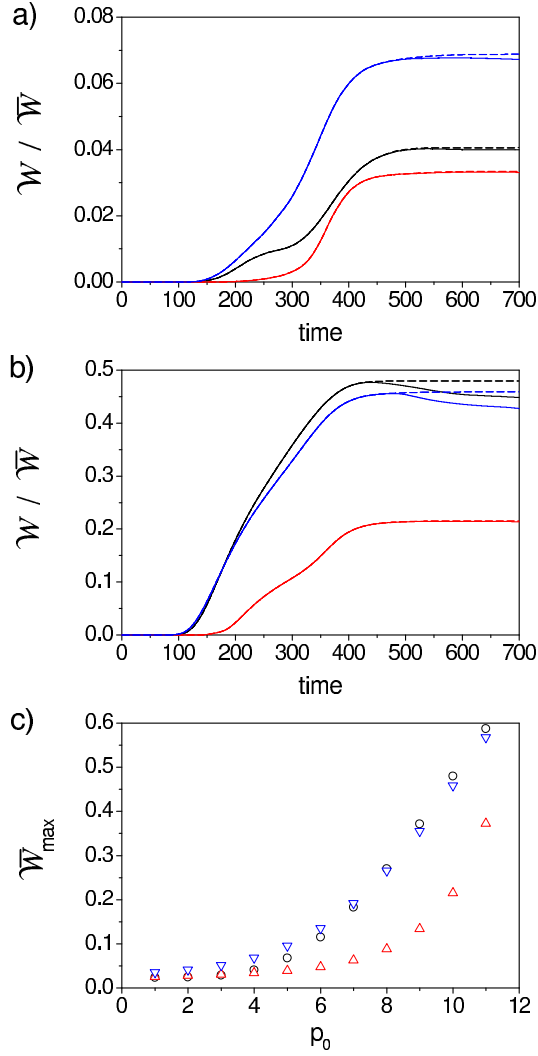}
 \caption{\label{fig3}
  Reaction probabilities $\mathcal{W}$ (solid curve) and
  $\bar{\mathcal{W}}$ (dashed curve) for: (a) $p_0 = 4$ and (b)
  $p_0 = 10$, and three different initial distributions: quantum
  trajectories distributed according to $\rho_0$ (black) and classical
  trajectories distributed according to $\rho_0$ (red) and $\rho_W$
  (blue).
  In panel (c), $\bar{\mathcal{W}}$ at $t \approx 700$ as a function
  of $p_0$: black circles denote $\bar{\mathcal{W}}_{\rm Bohm}$, red
  triangles, $\bar{\mathcal{W}}_{\rm Bohm}^{\rm cl}$, and blue
  inverted triangles, $\bar{\mathcal{W}}_{\rm Wigner}^{\rm cl}$.}
\end{figure}

In Figs.~\ref{fig3}(a) and \ref{fig3}(b), we have plotted $\mathcal{W}$
(solid line) and $\bar{\mathcal{W}}$ (dashed line) for $p_0 = 4$ and
$p_0 = 10$, respectively, and three different initial distributions:
quantum trajectories distributed according to $\rho_0$ (black) and
classical trajectories distributed according to $\rho_0$ (red) and
$\rho_W$ (blue).
Results from (\ref{eq-28}) have not been plotted since they are
identical to those obtained from (\ref{eq-29}).
In all cases, a total of $5 \times 10^4$ trajectories has been
considered  and a propagation up to $t = 700$, the asymptotic value
at which $\bar{\mathcal{W}}$ stabilizes.
As seen in Fig.~\ref{fig2}b, for $p_0 = 4$, $\bar{E}-\bar{\delta}$
(dashed curve) is well below $TS_2$ and $TS_1$.
Therefore, according to standard quantum mechanics, the actual dynamics
in Fig.~\ref{fig3}a should then proceed mainly via tunneling.
We clearly observe  that $\mathcal{W}_{\rm Bohm}$
is below $\mathcal{W}_{\rm Wigner}^{\rm cl}$ and closer to
$\mathcal{W}_{\rm Bohm}^{\rm cl}$.
This can be explained as follows.
The classical distributions explore many initial conditions, which
eventually may imply individual trajectory energies higher than the
transitions states and therefore lead to the formation of products.
This effect is more striking in the case of the Wigner distribution
than in the classical Bohmian one, where the initial momentum is fixed.
On the other hand, for Bohmian trajectories the dynamics is radically
different: for low $p_0$, the wave packet spreads faster than it
moves and, hence, the expected formation of ripples by interference is
going to hinder the passage of Bohmian trajectories to products.
Note that in Bohmian mechanics tunneling does not exist as it is
commonly understood in standard quantum mechanics, but as an effect
arising from the modification in time of the barrier due to the quantum
potential; a time-dependent, effective barrier, given by (\ref{eq-19}),
is what actually rules the trajectory dynamics.
For $p_0 = 10$, however, $\bar{E}-\bar{\delta}$ is above $TS_1$ in
Fig.~\ref{fig2}b, which suggests (quantum-mechanically) a larger
amount of products by direct passage rather than tunneling.
This will cause that the classical Wigner distribution and the (quantum)
Bohmian one will render much closer results, as it is observed in
Fig.~\ref{fig3}b, where $\mathcal{W}_{\rm Bohm}$ approaches
$\mathcal{W}_{\rm Wigner}^{\rm cl}$, but is far from
$\mathcal{W}_{\rm Bohm}^{\rm cl}$.
In this case, though tunneling can still be active, the direct passage
is going to control the dynamics in both cases, classical and
quantum-mechanical.
Observe that, in the second case, the translational motion is faster
than the spreading of the wave packet and, therefore, more trajectories
can be promoted to products before interference starts to influence the
trajectory dynamics.
Regarding $\bar{\mathcal{W}}$, we find a trend similar to $\mathcal{W}$,
but the difference between the asymptotic values of these magnitudes
increases with $p_0$, which is due to the larger amount of energy
available and therefore the increase of recrossings (for the classical
Bohmian distribution the effect is negligible, because of the less
energy available after fixing $p_0$).

In Fig.~\ref{fig3}c, the maximum amount of products,
$\bar{\mathcal{W}}$, is plotted as a function of $p_0$ for the same
distributions as above.
As can be noticed, the formation of products is more efficient
classically than quantum-mechanically until relatively large values
of $p_0$; indeed, both classical distributions provide a larger amount
of products for $p_0 \lesssim 4$.
As noticed in Fig.~\ref{fig2}b, $\bar{E}$ crosses $TS_2$ at $p_0
\approx 3.5$; in Fig.~\ref{fig3}c, after this $p_0$ value, the
Bohmian distribution provides larger reaction probabilities than its
classical counterpart, the difference between them increasing with
$p_0$.
As $p_0$ is further increased, the Bohmian and Wigner distributions
start to get closer, though the latter gives rise to a larger
probability until $p_0 \sim 8$-9, where the situation is reversed.
Again, this coincides with $\bar{E}$ crossing a transition state,
this time the one connecting the pre-equilibrium with products.

Summarizing, though an ensemble magnitude may have very similar values
classically and quantum-mechanically, the underlying dynamics can be
radically different, which arises from both the effects of the quantum
potential and the mathematical structure of the corresponding motion
equations (see Sec.~\ref{sec2}): in Bohmian mechanics the momentum
readjusts at each instant in order to satisfy the statistical
requirements of quantum mechanics, while in classical mechanics
positions and momenta are not coupled.
Next, we analyze in more detail the underlying dynamics in terms
of common elements within the generalized HJ framework: RPs and
classical/quantum trajectories.

%%%%%%%%%%%%%%%%%%%%%%%%%%%%%%%%%%%%%%%%%%%%%%%%%%%%%%%%%%%%%%%%%%%%%%%

\subsection{Quantum-classical trajectory analysis}
\label{3.3}

\begin{figure}
\includegraphics[width=7cm]{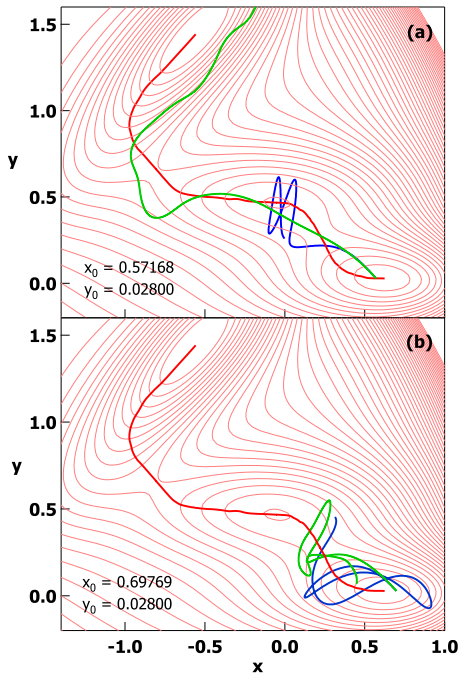}
 \caption{\label{fig4}
 Classical (blue) and quantum (green) trajectories representative of
 the dynamics on a M\"uller-Brown PES (in red, the RP): (a) the quantum
 trajectory is reactive through two-dimensional tunneling, while its
 classical counterpart is inelastic; (b) both trajectories are
 inelastic. The classical and quantum trajectories are launched
 from the initial positions indicated at the lower left corner of each
 respective panel and with the same initial momentum, $p_0 = 9$.}
\end{figure}

As seen in Fig.~\ref{fig2}b, values of $p_0$ between 8 and 10 are
critical, since the dynamics is developed around the $TS_1$ from the
pre-equilibrium region.
Thus, in Fig.~\ref{fig4}, two pairs of trajectories for $p_0 = 9$ have
been purposely chosen.
In each panel, both members of the pair start exactly with the same
initial conditions.
However, in (a) the quantum trajectory (green) is reactive through
two-dimensional tunneling, while its classical partner (blue) is
inelastic, whereas in (b) both trajectories are inelastic.
The different and richer behaviors displayed by quantum trajectories
are due to their motion being ruled by the quantum potential (apart
from the PES), which is lacking in classical dynamics.

\begin{figure}
\includegraphics[width=9cm]{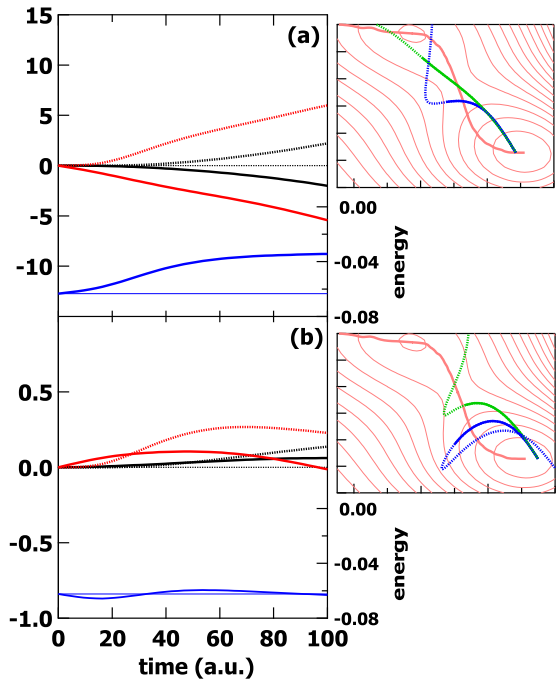}
\caption{\label{fig5}
 Quantum-classical differences for the pairs of trajectories displayed
 in Figs.~\ref{fig4}(a) and \ref{fig4}(b): $\Delta x (t)$ (black
 solid), $\Delta y (t)$ (black dashed), $\Delta p_x (t)$ (red solid)
 and $\Delta p_y (t)$ (red dashed).
 The total energy (\ref{eq-26}) is also displayed (the scale is on
 the right side of each panel), with blue thinner/thicker line for
 classical/quantum trajectories (the classical energy is constant
 with time).
 In the aside plots, solid lines indicate the portion of the
 trajectories up to the time when one member of the pair approaches
 the first saddle point ($t \approx 100$~a.u.).}
\end{figure}

In order to better understand these behaviors, we consider differences
$\Delta A(t) = A_q(t) - A_{cl}(t)$, where $A_q(t)$ and $A_{cl}(t)$
denote the quantum magnitude and its classical counterpart, respectively.
Thus, in Fig.~\ref{fig5}, we show the differences in positions
($\Delta x$ and $\Delta y$) and momenta ($\Delta p_x$ and $\Delta p_y$)
for the pairs of trajectories displayed in Fig.~\ref{fig4} up to the
time when one member of the pair approaches the first saddle point
($t \approx 100$~a.u.).
Moreover, the total energy, computed from the classical-like expression
\begin{equation}
 E(t) = \frac{p_x^2(t)}{2m} + \frac{p_y^2(t)}{2m} + V(x(t),y(t)) ,
 \label{eq-26}
\end{equation}
is also plotted along each classical (thinner)/quantum (thicker)
trajectory (blue lines) taking into account the energy scale in the
right vertical axes of both panels of the corresponding figure.
At $t=0$, (\ref{eq-26}) is between $V_{TS_2}$ and $V_{TS_1}$ (see
Fig.~\ref{fig2}b) for both classical trajectories and, therefore,
none of them will be able to overcome the barrier leading to products
at any subsequent time since, classically, the energy conserves.
Hence, they will remain wandering within the reactants and
pre-equilibrium regions.
Regarding the quantum dynamics, the trajectory in Fig.~\ref{fig4}a
basically follows the RP, thus reaching the products region.
When compared with its classical counterpart, we note that,
at the earlier stages of its time evolution, the separation of
both types of trajectories is increasing.
This fact suggests that the quantum trajectory is visiting a wider
region of phase space than the corresponding classical one (termed as
{\it dynamical tunneling} by Heller \cite{heller}), even before
reaching the potential barrier (barrier tunneling).
Thus we could say that we have  a double contribution to tunneling.
The main contribution to the reaction probability will come from the
dynamical tunneling since, in two or higher dimensions, the trajectory
can skip the barrier very easily by exploring new regions of phase
space.
On the contrary, the quantum trajectory of Fig.~\ref{fig4}b is
inelastic because there is no route to products mediated by tunneling.
Arguing in topological terms, this trajectory is launched pointing
towards a part of the PES with a stronger gradient (i.e., a more
repulsive wall); the trajectory is not capable then to overcome the
increasing bending and go backwards, so it remains trapped at the
pre-equilibrium region.
This dynamical picture of tunneling in two dimensions thus seems to be
more demanding than in one dimension, since not only energy criteria,
but also the initial momentum orientation should be taken into account.

Energetically, quantum dynamics look so different from classical ones
(for instance, we may observe reactivity quantum-mechanically, but not
classically, as mentioned above) because (\ref{eq-26}) does not
conserve along quantum trajectories.
This can be appreciated in Fig.~\ref{fig5}, when comparing the value of
this magnitude for the two quantum trajectories along time.
In Fig.~\ref{fig5}a, there is a relatively fast initial expansion
or `boost' \cite{wyatt1,wyatt2} which leads to an increase of the
energy and therefore enhances the possibility of tunneling though
the initial energy was smaller than the barrier height.
On the contrary, the second trajectory does not benefit from this
expansion (its energy remains nearly conserved) and, therefore, no
tunneling is expected.

%%%%%%%%%%%%%%%%%%%%%%%%%%%%%%%%%%%%%%%%%%%%%%%%%%%%%%%%%%%%%%%%%%%%%%%

\subsection{Statics-dynamics analysis}
\label{3.4}

\begin{figure}
\includegraphics[width=7cm]{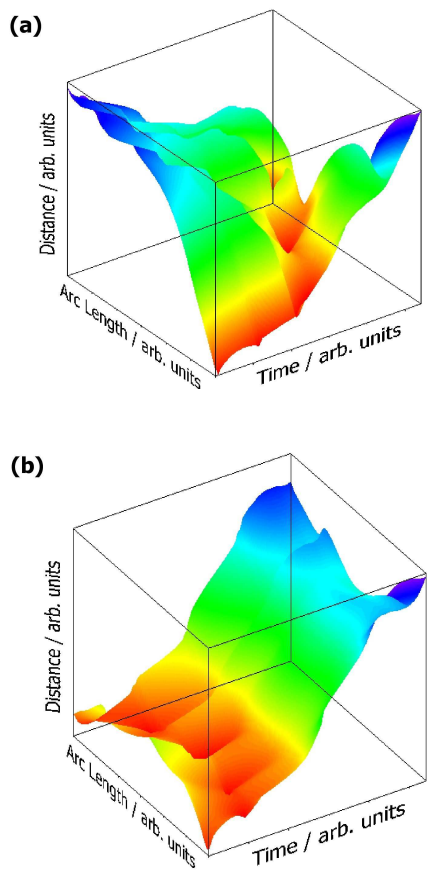}
 \caption{\label{fig6}
 In panels (a) and (b), Carath\'eodory plots associated with
 the quantum trajectories displayed in Figs.~\ref{fig4}(a) and
 \ref{fig4}(b), respectively.
 The transition from red to blue indicates an increasing distance
 between the RP and the respective quantum trajectory.}
\end{figure}

Finally, we would like to provide some clues aimed to clarify the
long-standing static-dynamic controversy, i.e., whether statics suffices
to establish reaction mechanisms or, on the contrary, true reaction
mechanisms are very different from the information provided by statics
and, therefore, dynamics is compulsory.
A direct approach to tackle this problem consists of studying the
differences in configuration space between the RP and classical/quantum
trajectories.
In principle, any static-dynamic comparison might be hampered by the
different physical meaning of the independent parameter characterizing
each curve as well as the dimensionality of the problem.
In order to avoid these difficulties, though at the expense of losing
some dynamical details, one can consider projection schemes.
For instance, trying to check the accuracy of the RP formalism,
Taketsugu and Gordon \cite{gordon} tackled such a problem using two
distances, the perpendicular and the distance of closest approach.
We propose a generalization of this approach by computing the whole
set of distances between the RP and a given trajectory.
This generates a matrix with elements
\begin{equation}
 C_{ij} = |{\bf x}(t_i) - {\bf x}_{\rm RP}(s_j)|^2 ,
 \label{eq-25}
\end{equation}
where the $(i,j)$-element gives the distance between a (classical
or quantum) trajectory, ${\bf x}(t_i)$, at a time $t_i$, and the RP
${\bf x}_{\rm RP}(s_j)$, at a value $s_j$ of the reaction
coordinate (of course, the same division in segments is assumed for
the trajectory and the RP), and whose visual display gives rise to
the so-called {\it Carath\'eodory plot} \cite{cara}.
These plots have the property that they are always three-dimensional
surfaces regardless of the number of degrees of freedom of the problem
because they are based on the definition of distance (\ref{eq-25}).
In Fig.~\ref{fig6}, we present the Carath\'eodory plots associated
with the quantum trajectories displayed in Figs.~\ref{fig4}(a) and
\ref{fig4}(b).
The quantum trajectory of Fig.~\ref{fig4}a is able to tunnel across
the highest TS and reach the products region keeping its kinetic energy
considerably lower than that of classical reactive trajectories.
Hence, it remains reasonably close to the RP rather than oscillating
around it (as it might be expected classically), this leading to a
Carath\'eodory plot which displays a valley essentially along its
diagonal (see Fig.~\ref{fig6}a).
On the contrary, the quantum trajectory in Fig.~\ref{fig4}b is
inelastic and, therefore, departs remarkably from the RP, this giving
rise to a minimum valley out of the diagonal (see Fig.~\ref{fig6}b).

%%%%%%%%%%%%%%%%%%%%%%%%%%%%%%%%%%%%%%%%%%%%%%%%%%%%%%%%%%%%%%%%%%%%%%%

\section{Conclusions}
\label{sec4}

Here, we have shown the guidelines to carry out analyses of molecular
reactions within a generalized HJ framework, which allows to encompass
both topology and classical/quantum dynamics: static RPs arise when a
variational kernel is associated with a PES gradient, whereas
classical/quantum trajectories are obtained when it corresponds
to properly defined classical/quantum Lagrangians.
Specifically, the application of this scheme to the topology and
dynamics in a M\"uller-Brown PES has shown outstanding features, such
as a better understanding of the discrepancies found between classical
and quantum statistical results.
For instance, reaction probabilities have been computed with Bohmian
and Wigner distributions of initial conditions.
As seen, though both rates look pretty close for some values of $p_0$,
their corresponding underlying dynamics are dramatically different
because of the quantum-mechanical coupling between individual
trajectories and their distribution.
This leads to dynamics totally different from their classical analogs
and, therefore, to effects not observable classically, such as dynamical
tunneling.
It is also worth stressing that computing $\bar{\mathcal{W}}$ and
displaying its asymptotic value as a function of $p_0$ (or another
parameter characterizing the initial state), can be of interest from an
experimental viewpoint, for it constitutes a simple method to determine
the maximum amount of products formed during a reaction.
Since $\bar{\mathcal{W}}_{\rm max}$ provides an upper bound for the
reaction probability, it could be used to establish some control
mechanism on the reaction based on modifying one parameter in the
initial state preparation (here, $p_0$, which could be varied by
using laser pumping techniques, for instance).
On the other hand, quantum-classical differences, obtained from a
one-to-one comparison of classical and quantum trajectories with the
same initial conditions, give evidence of dynamical tunneling at
early time evolution stages, this contribution being mainly the
responsible for the occurrence of reaction at certain initial
conditions. Barrier tunneling is expected to be very small.
Finally, the construction of Carath\'eodory plots has been shown to
be an interesting tool to analyze static-dynamic differences (i.e.,
between RPs and classical/quantum trajectories).
Depending on whether the trajectory is reactive or not, these plots
show distinct topographical shapes: tunneling reactive trajectories
lead to diagonally dominated valleys, whereas inelastic, non-reactive
ones lead to a much more involved shape.
Furthermore, since these plots do not depend on the problem
dimensionality, they could result advantageous to analyze processes
involving large molecules.

%%%%%%%%%%%%%%%%%%%%%%%%%%%%%%%%%%%%%%%%%%%%%%%%%%%%%%%%%%%%%%%%%%%%%%%

\acknowledgements

This work has been supported by the Ministerio de Ciencia
e Innovaci\'on (Spain) under Projects FIS2007-62006 and
CTQ2008-02856/BQU, and Generalitat de Catalunya under Projects
2005SGR-00111 and 2005SGR-00175.
A.\ S.\ Sanz would also like to acknowledge the Consejo Superior de
Investigaciones Cient\'{\i}ficas for a JAE-Doc Contract.

%%%%%%%%%%%%%%%%%%%%%%%%%%%%%%%%%%%%%%%%%%%%%%%%%%%%%%%%%%%%%%%%%%%%%%%

\end{document}